\documentclass[usenatbib]{mn2e} 
\usepackage{longtable} 
\usepackage{graphicx}

\newcommand{\kms}{\,{\rm km}\,{\rm s}$^{-1}$}

\def\aj{AJ}
\def\araa{ARA\&A}
\def\apj{ApJ}
\def\apjs{ApJS}
\def\aa{A\&A}
\def\aas{A\&AS}
\def\bac{Bull. astr. Inst. Czechosl.}
\def\mnras{MNRAS}
\def\pasp{PASP}
\def\fcp{Fund.~Cosmic~Phys.}

\title[Study of open cluster NGC 637 and NGC 957]
      {Optical and near-IR photometric study of the open cluster 
       NGC 637 and NGC 957 
      }

\author[Yadav et al.]
   {R. K. S. Yadav$^{1}$\thanks{E-mail: rkant@aries.ernet.in}, 
    Brijesh Kumar$^{1,2}$, A. Subramaniam$^{3}$, Ram Sagar$^{1}$ and 
    \newauthor Blesson Mathew$^{3}$ \\ 
    $^{1}$ Aryabhatta Research Institute of Observational Sciences, 
           Manora Peak, Nainital 263129, India\\
    $^{2}$ Departamento de F\'isica, Universidad de 
           Concepci\'on, Casilla 160-C, Concepci\'on, Chile\\
    $^{3}$Indian Institute of Astrophysics, Bangalore 560 034, India  
   }

\begin{document}

\date{\today}


\maketitle

\label{firstpage}

\begin{abstract}
  We present $UBVRI$ CCD photometry in the region of the open clusters
  NGC 637 and NGC 957. The radii are found to be 4\farcm2 and 4\farcm3. Their
  reddenings $E(B-V)$ are $0.64\pm0.05$ mag and $0.71\pm0.05$ mag and their
  distances, from main sequence fitting are $2.5\pm0.2$ kpc and $2.2\pm0.2$ kpc.
  Comparison with $Z=0.02$ isochrones leads to an age of $10\pm5$ Myr for both
  clusters. Combining our photometry with 2MASS JHK shows the reddening law
  in these directions to be normal. Mass function slopes of $x=1.65\pm0.20$
  and $1.31\pm0.50$ are derived for the clusters, both of which are found to
  be dynamically relaxed. Spectral and photometric characteristics of three 
  Be stars, two in NGC 957 and one (newly discovered) in NGC 637 indicate 
  them to be of Classical Be type." 

\end{abstract}



\section{Introduction} \label{sec:intro}

Open clusters are excellent targets to understand issues related to Galactic 
structure, chemical composition, stellar population, dynamical evolution and 
star formation processes in the Galaxy. The importance of photometric studies 
of star clusters lie in their colour-magnitude diagrams (CMDs) which allow us 
to estimate cluster's fundamental parameters such as age, 
distance and reddening. In this study, we consider two young open cluster, 
NGC 637 and NGC 957, situated in perseus arm of Milky Way to estimate 
their fundamental parameters and mass function. 

{\bf NGC 637} = C0139$+$637 ($\alpha_{2000} = 01^{h}43^{m}04^{s}$, 
$\delta_{2000}$=$+$64\degr02\arcmin24\arcsec; $l$ = 128\fdg55, $b$=1\fdg73) is 
classified as a Trumpler class I2m by \citet{ruprecht66}. Using $RGU$ 
photographic data, \citet{grubissich75} found the cluster to be located 
in the Perseus arm at a distance of 2.4 kpc. \citet{huestamendia91} used 
the $UBV$ photoelectic photometry and derived a colour excess 
of $E(B-V)=0.66$ mag and a distance of 2.5 kpc. Recently, 
\citet{phelps94} studied this cluster using $UBV$ CCD photometry and 
estimated a colour excess of $E(B-V)=0.49\pm0.03$ mag. Further, using CMD and 
assuming a total-to-selective absorption ($R_V$) of 3.1, they derived a 
distance of 2.75 kpc. 

{\bf NGC 957} = C0230$+$573 ($\alpha_{2000} = 02^{h}33^{m}21^{s}$,
$\delta_{2000}$=57\degr33\arcmin36\arcsec; $l$=136\fdg28, $b$=$-$2\fdg65) 
is located in the Perseus spiral arm of our Galaxy and according to 
\citet{ruprecht66}, it is of type III 2p. This cluster was studied by 
\citet{gimenez80} using $RGU$ photographic photometry and they found 
that this is an young star cluster situated at a distance of 1.8 kpc. 
This cluster was further investigated by \citet{gerasimenko91} using 
Johnson $UBV$ photoelectric and photographic photometry. He derived 
reddening of $E(B-V)=0.90$ mag, distance of 2.1 kpc and age of about 4 Myr. 

In the present study, we provide deep $UBVRI$ CCD photometry of NGC 637 and 
NGC 957, and re-visit their basic parameters and estimate their luminosity 
and mass functions. Both the clusters are poorly studied in the literature. 
For NGC 957, CCD photometric observations is presented for 
the first time alongwith spectroscopic observations of a few brightest 
members of both the clusters. 
  
The paper is organized as follows. We present the observations and data 
reductions in \S\ref{sec:obs}, while in \S\ref{sec:ana}, we describe
the data analysis. \S\ref{sec:mfu} and \S\ref{sec:mse} describe about 
the mass function and mass segregation. Finally, \S\ref{sec:con} contains 
the conclusions of our study.

  \begin{figure*}
    \centering
    \includegraphics[width=18cm, height=18cm]{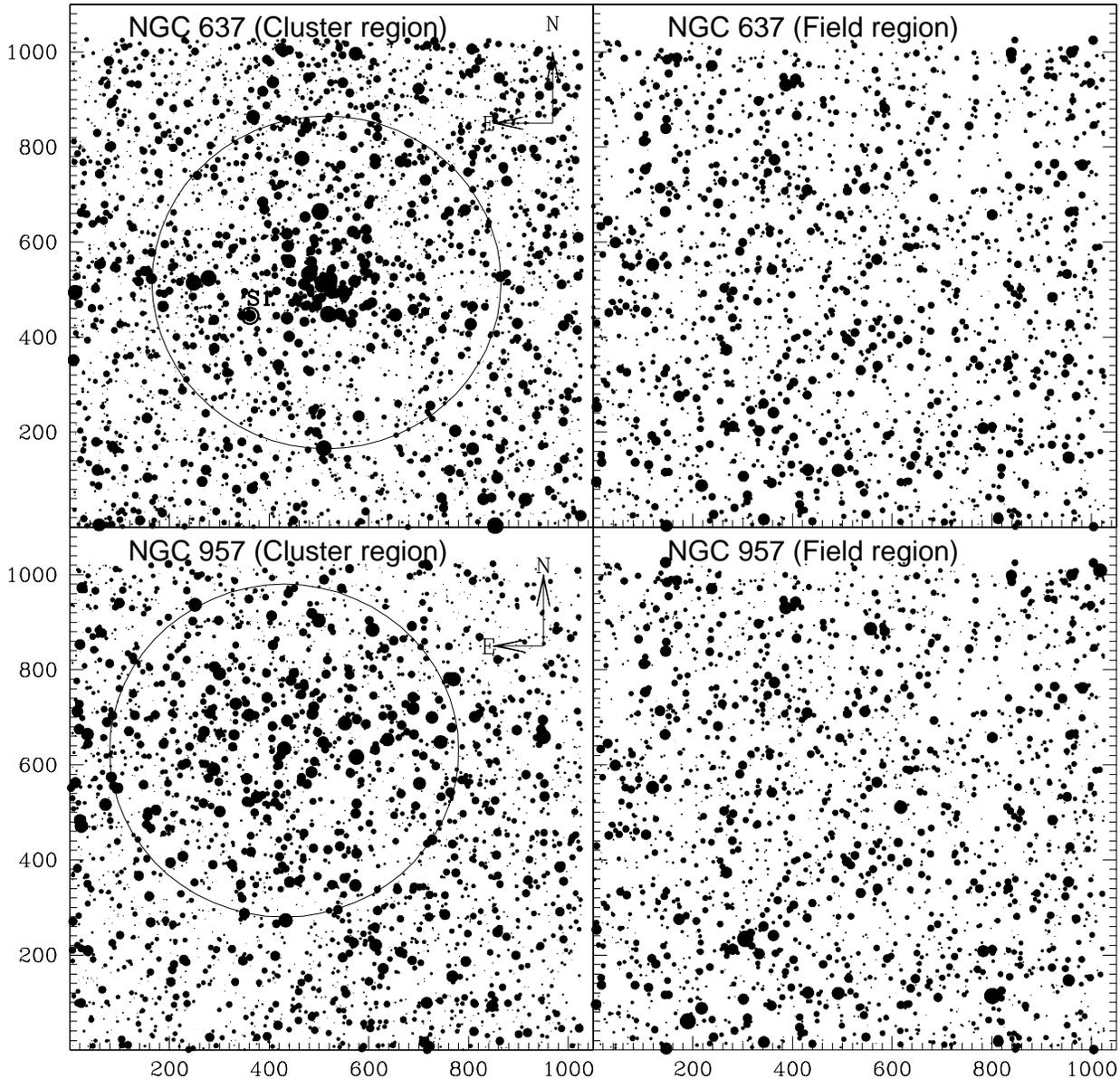}
    \caption{Finding chart of the stars in the field of NGC 637 and NGC 957. 
The (X, Y) coordinates are in pixel units corresponding to 0$^{\prime\prime}$.72 
on the sky. Direction is indicated in the map. Filled circles of different 
sizes represent brightness of the stars. Smallest size denotes stars of 
$V$$\sim$20 mag. Open circle represent the cluster size. S1 star in the region 
of NGC 637 is emission type star.}
  \label{fig:chart}
  \end{figure*}

\section{observations and data reduction} \label{sec:obs}

\subsection {Photometric observations}
Johnson $UBV$ and Cousins $RI$ CCD photometric observations were obtained 
using 104-cm Sampurnanand telescope at Nainital on 13 November 2004 
for NGC 637 and on 28 September 2006 for NGC 957. Data were recorded 
using a liquid nitrogen cooled 2k$\times$2k CCD of 24$\mu$m square size 
pixel, which results in a scale of 0\farcs36 per pixel and a square field 
of 12\farcm7 in the sky on a side. The CCD has a gain of 10 electrons per 
analog-to-digital unit and a read out noise of 5.3 electrons. To improve 
the signal-to-noise ratio, all the observations were taken in $2\times2$ 
binning mode. Log of observations are given in Table 1. In $V$ and $I$ 
filters, we also observed field regions situated about 15\arcmin north from 
the cluster center of both NGC 637 and NGC 957. Identification maps of 
cluster and field regions are shown in Fig. 1.
\begin{table}
\centering
\scriptsize
\caption{Log of observations, with dates and exposure times for each passband.
$N$ denotes the number of stars measured in different passband.}
\begin{tabular}{cccc}
\hline\hline
Band  &Exposure Time &Date&$N$\\
&(in seconds)   & &\\
\hline\hline
&&NGC 637(Cluster region)&\\
$U$&1800$\times$2, 300$\times$1&2004 November 13/14&1057\\
$B$&1200$\times$3, 300$\times$1&2004 November 13/14\\
   & 120$\times$1; 10$\times$1 &2004 November 13/14&2392\\
$V$&900$\times$3, 120$\times$1; 10$\times$1 &2004 November 13/14&3261\\
$R$&480$\times$3, 60$\times$2; 5$\times$1 &2004 November 13/14&3219\\
$I$&300$\times$3, 60$\times$1; 5$\times$1 &2004 November 13/14&3151\\
&&NGC 637(Field region)&\\
$V$&180$\times$2&2004 November 14/15&1286\\
$I$&180$\times$2&2004 November 14/15&1266\\
\hline
&&NGC 957&\\
$U$&1800$\times$2, 300$\times$1&2006 September 28/29&915\\
$B$&1200$\times$2, 300$\times$2&2006 September 28/29&2087\\
$V$&900$\times$2, 180$\times$2&2006 September 28/29&2855\\
$R$&600$\times$2, 120$\times$2&2006 September 28/29&2819\\
$I$&300$\times$3, 60$\times$2&2006 September 28/29&2707\\
&&NGC 957(Field region)&\\
$V$&180$\times$2&2004 November 14/15&1284\\
$I$&180$\times$2&2004 November 14/15&1269\\
\hline
\end{tabular}
\label{log}
\end{table}

A number of bias and flat-field frames were taken during the observations. 
Flat-field exposures were taken of the twilight sky in each filter. Bias 
and flat field corrections were performed using the standard 
IRAF\footnote{IRAF is distributed by the National Optical Astronomical 
Observatory which are operated by the Association of Universities for Research 
in Astronomy, under contract with the National Science Foundation} tasks. 
Subsequent data reduction and analysis were done using the DAOPHOT software 
\citep{stetson87, stetson92}. We performed the profile fitting photometry
using quadratically varying point spread function. For each filter, the 
stars have been aligned to that of a reference frame with largest exposure
time and an average instrumental magnitude weighted by the photometric error 
was derived. A final catalogue was created of stellar objects identified in 
at least two filters, with shape-defining parameter 
$-2 \le {\it sharpness} \le 2$ and  and the goodness-of-fit 
estimator $\chi \le 5$.

  \begin{figure}
    \centering
    \includegraphics[width=9cm]{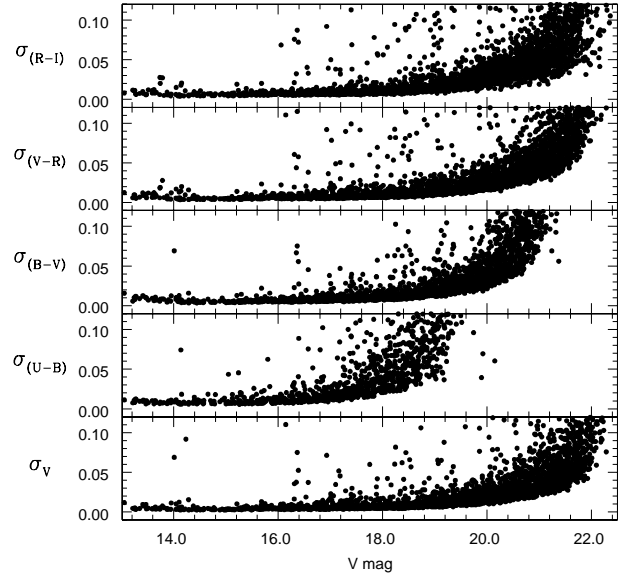}
    \caption{The photometric errors (in magnitude) corresponding to the 
       brightness measurement at $V$, $U-B$, $B-V$, $V-R$ and $V-I$, are 
       plotted against the $V$-band brightness. Errors on the y-axis represent 
       the internal errors as estimated by DAOPHOT routine.} 
    \label{fig:error}
  \end{figure}

\begin{table}
\caption{Comparison of our photometry with others for NGC 637 and NGC 957. The difference 
($\Delta$) is always in the sense present minus comparison data. The mean along with their 
standard deviations in magnitude are based on N stars. Few deviated points are not included 
in the average determination.}
\tiny
  \begin{tabular}{cccc}
  \hline
  $V$ range&$<\Delta$$V>$&$<\Delta(B-V)$$>$&$<\Delta(U-B)>$\\
  &Mean$\pm$$\sigma$(N)&Mean$\pm$$\sigma$(N)&Mean$\pm$$\sigma$(N)\\
  \hline
&&&\\
               & {\bf NGC 637}&&\\
&&&\\
               & Comparison with&&\\
               & Huestamendia et al. (1991)&&\\
  10.0 $-$ 11.0&0.01$\pm$0.01(2)&0.01$\pm$0.01(2)&0.02$\pm$0.02(2)\\
  12.0 $-$ 13.0&0.03$\pm$0.05(3)&0.01$\pm$0.01(3)&0.03$\pm$0.02(3)\\
  13.0 $-$ 14.0&0.05$\pm$0.04(8)&0.01$\pm$0.02(8)&0.03$\pm$0.04(8)\\
  14.0 $-$ 15.0&0.02$\pm$0.02(7)&0.01$\pm$0.01(7)&0.04$\pm$0.02(4)\\

&&&\\
               & Comparison with &&\\
               & Phelps \& Janes (1994)&&\\
&&&\\
  10.0 $-$ 11.0&0.05$\pm$0.06(4)  &&\\
  12.0 $-$ 13.0&0.05$\pm$0.04(2)  &0.03$\pm$0.01(3)&0.13$\pm$0.03(3)\\
  13.0 $-$ 14.0&0.08$\pm$0.06(12) &0.02$\pm$0.04(14)&0.13$\pm$0.06(14)\\
  14.0 $-$ 15.0&0.06$\pm$0.07(13) &0.03$\pm$0.03(15)&0.14$\pm$0.08(14)\\
  15.0 $-$ 16.0&0.04$\pm$0.04(12) &0.01$\pm$0.05(21)&0.13$\pm$0.05(15)\\
  16.0 $-$ 17.0&0.03$\pm$0.04(44) &0.02$\pm$0.04(45)&0.12$\pm$0.06(27)\\
  17.0 $-$ 18.0&0.03$\pm$0.05(61) &0.01$\pm$0.05(65)&0.10$\pm$0.07(42)\\
  18.0 $-$ 19.0&0.03$\pm$0.08(67) &0.01$\pm$0.06(73)&0.11$\pm$0.08(27)\\
  19.0 $-$ 20.0&0.03$\pm$0.07(102)&0.01$\pm$0.09(96)&\\
&&&\\
& {\bf NGC 957}&&\\
&&&\\
&Comp. with \\
&Gimenez \& Garcia-pelayo\\
&(1980)\\
&&&\\
  12.0 $-$ 13.0&$-$0.06$\pm$0.09(5) &$-$0.03$\pm$0.05(5)&0.03$\pm$0.05(5)\\
  13.0 $-$ 14.0&$-$0.04$\pm$0.07(6) &$-$0.03$\pm$0.06(6)&0.03$\pm$0.06(6)\\
  14.0 $-$ 15.0&$-$0.03$\pm$0.06(3) &$-$0.03$\pm$0.06(3)&0.03$\pm$0.06(3)\\
  \hline

\end{tabular}
\label{comput_dif}
\end{table}


  \begin{figure}
   \centering
   \includegraphics[width=8cm]{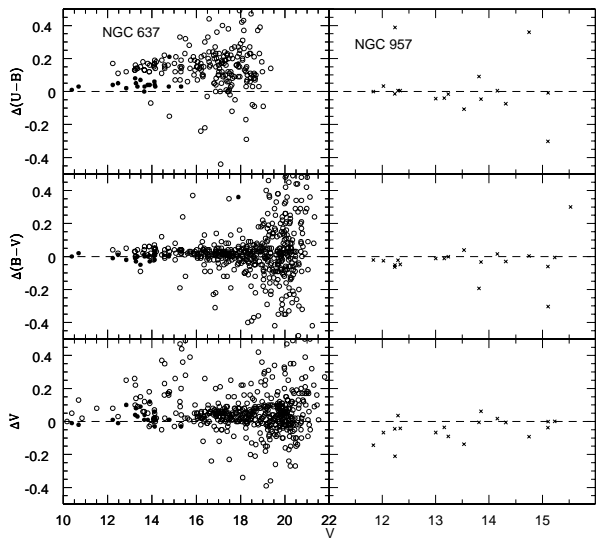}
   \caption{Comparison of present CCD $UBV$ photometry with that available in 
         the literature. For NGC 637, the solid ($\bullet$) circles represent 
         data from \citet{huestamendia91} while the open circles ($\circ$) 
         come from \citet{phelps94}. For NGC 957, the crosses ($\times$) 
         represent comparison with photographic data taken from 
         \citet{gerasimenko91}.}
   \label{fig:litmch}
  \end{figure}

For translating the instrumental magnitude to the standard magnitude, we 
observed the standard field PG 0231$+$051 \citep{landolt92} on 20 
November 2004 several times during the night to determine the Earth's 
atmospheric extinction coefficients, color coefficients and the zero points
of the telescope and CCD system. The standard stars used in the calibrations 
have brightness range of $12.77 \le V \le 16.11$ and a color range of 
$-0.329 < (B-V) < 1.448$. The transformation equations derived using linear 
least square regression are as follows:

\begin{center}
   $u=U+6.92\pm0.01-(0.06\pm0.01)(U-B)+(0.54\pm0.02)X$
   $b=B+4.66\pm0.01-(0.02\pm0.01)(B-V)+(0.24\pm0.01)X$
   $v=V+4.22\pm0.01-(0.01\pm0.01)(B-V)+(0.14\pm0.01)X$
   $r=R+4.13\pm0.01-(0.00\pm0.01)(V-R)+(0.10\pm0.01)X$
   $i=I+4.60\pm0.01-(0.02\pm0.02)(R-I)+(0.06\pm0.01)X$
\end{center}

\noindent 
where $u, b, v, r$ and $i$ are the aperture instrumental magnitudes
and $U, B, V, R$ and $I$ are the standard magnitudes and $X$ is 
the airmass. Zero-point and colour-coefficient errors are $\sim$ 0.01 mag. 

The internal errors, as derived from DAOPHOT, in magnitude and colour 
are plotted against $V$ magnitude in Fig.~\ref{fig:error}. This figure 
shows that photometric error is $\le$ 0.01 mag at $V\sim19$ mag while  
error in different colours is $\le$ 0.02 mag. The final photometric data 
are available in electronic form at the WEBDA 
site \footnote{\it http://obswww.unige.ch/webda/} and it can also 
be obtained from the authors.

We cross identified our objects with that present in the literature.
Fig.~\ref{fig:litmch} shows the plots of difference $\Delta$ in $V$, $(B-V)$ 
and $(U-B)$ with $V$ magnitude. Solid and open circles represent the 
difference with photoelectric and CCD data. The average difference 
alongwith their standard deviation are listed in Table 2. The comparison of 
present and photoelectric data in $V$ magnitude and $(B-V)$, $(U-B)$ colours 
in NGC 637 do not show any significant difference with stellar magnitudes. 
Comparison of NGC 957 CCD data with photographic data also do not show any 
significant difference. So, we conclude that the present CCD photometry is 
in good agreement with photoelectic and photographic data for NGC 637 and 
NGC 957 respectively. On the other hand for NGC 637, comparison of CCD data 
with \citet{phelps94} show a zero-point offset of $\sim$ 0.04 mag in $V$ 
and $\sim$ 0.12 mag in $(U-B)$ while there is no noticeable difference seen 
in $(B-V)$. A possible cause of the zero-point offset may be due to the 
difference in dates of observation for \citet{landolt92} standards and the 
target cluster fields. Since the zero-points depend on the factors, such as 
ambient temperature and Earth's atmospheric conditions, these coefficients 
change from night-to-night. A typical nightly variation in zero-points 
is found to be $\sim$ 0.08 mag in $U$ and $\sim$ 0.05 mag in $B, V, R$ 
and $I$ passband. In addition to this $\sim$ 0.02 mag nightly variation 
is found in the extinction coefficients.

\begin{table}
\scriptsize
\centering
\caption{Log of slitless and slit spectroscopic observations, with dates and exposure times.}

\hspace{2cm}

\begin{tabular}{lccc}
\hline\hline
Object&Band  &Exposure Time &Date\\
&&(s)   & \\
\hline\hline
NGC 637&R&5&2004 January 28 \\
&R+Gr5&60&2004 January 28\\
&R+Gr5&600&2004 January 28\\
NGC 957(East)&R&5&2004 January 29\\
&R+Gr5&60&2004 January 29\\
&R+Gr5&600&2004 January 29\\
NGC 957(West)&R&5&2004 January 29\\
&R+Gr5&60&2004 January 29\\
&R+Gr5&600&2004 January 29\\
NGC 957(West)&R&5&2004 January 29\\
&R+Gr5&60&2004 January 29\\
NGC 637(1) & Grism 7/167l & 900& 2006 September 30\\
           &             Grism 8/167l & 900&2006 September 30\\
NGC 957(1) & Grism 7/167l & 300& 2006 September 30\\
           &             Grism 8/167l & 300& 2006 September 30\\
NGC 957(2) & Grism 7/167l & 600& 2006 September 30\\
           &             Grism 8/167l & 600& 2006 September 30\\
\hline
\end{tabular}
\label{log_spec}
\end{table}

\subsection{Spectroscopic observations}

We obtained slitless spectra of stars in the cluster regions using Himalayan 
Faint Object Spectrograph and Camera mounted with the 2-m Himalayan 
Chandra Telescope at Hanle, India\footnote{http://www.iiap.res.in}. Grism 5 
($\lambda \sim 5200-10300$ \AA{}, $\lambda/\Delta \lambda \sim 870$) was used as 
dispersing element in combination with $R$ filter 
($\lambda \sim 7000$ \AA{}, $\Delta \lambda \sim 2200$ \AA{}).
The slitless spectra were recorded on a 2k$\times$2k CCD with 15$\mu$m pixel 
size and with an image scale of 0\farcs297 per pixel, which covered a total 
area of around $10\arcmin \times10\arcmin$. A typical spectral resolution 
of 6 \AA{} is achieved at H$_{\alpha}$. We list the log of observations in 
Table 3. We obtained the slit spectra of the three identified 
emission line stars on 30 September 2006 using Grism 7 ($3800-6800$\AA{}) 
and 167 $\mu$m slit combination in the blue region which gives an effective 
resolution of 1330. The spectra in the red region were obtained on the same 
night using Grism 8 ($5800-8350$ \AA{}) and 167 $\mu$m slit, which gives 
an effective resolution of 2190. The Gaussian FWHM seeing was 
about $1\farcs5$ ($\sim5$ pixel).
  \begin{figure}
    \centering
    \hbox{
    \includegraphics[width=9cm]{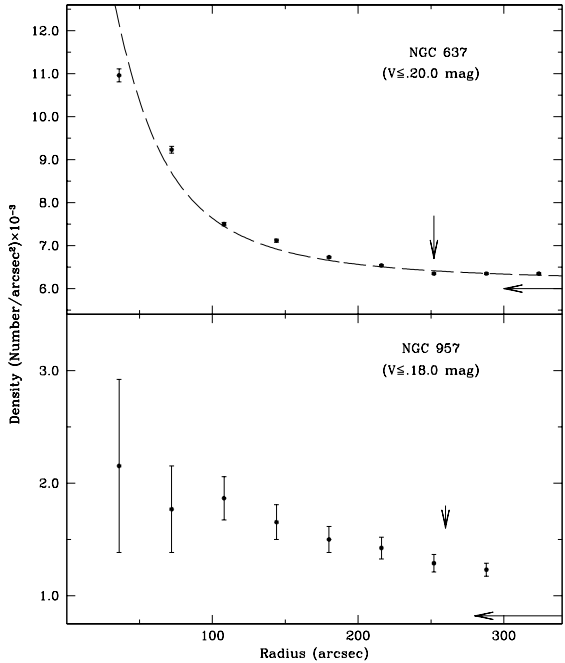}
    }
    \caption{Surface density distribution of stars in the field of the cluster 
      NGC 637 and NGC 957. Errors are determined from sampling 
      statistics(=$1/\sqrt{N}$ where $N$ is the number of stars used 
      in the density estimation at that point). Vertical and horizontal arrows 
      represent the radius of the clusters and field star density 
      respectively.} 
    \label{fig:dens}
  \end{figure}

\section{Analysis of the data} \label{sec:ana}

\subsection{Cluster radius and radial stellar surface density}

We perform star counts in concentric rings around an estimated centre of 
the cluster, and then divide it by their respective areas. The cluster center 
is estimated iteratively by calculating average X and Y position of the stars 
within 400 pixels from an eye estimated center, until they converged to a 
constant value. The cluster center in pixel units derived by this method 
are (515, 515) for NGC 637 and (430, 630) for NGC 957.
The cluster center in celestial coordinate are 
($\alpha_{2000} = 01^{h}43^{m}04^{s}.2$,
$\delta_{2000}=64\degr02\arcmin23\farcs5$) for NGC 637 and 
($\alpha_{2000} = 02^{h}33^{m}23^{s}.9$,
$\delta_{2000}=57\degr33\arcmin29\farcs9$ for NGC 957. The derived 
center of both the clusters are close to the one given in WEBDA 
(\S\ref{sec:intro}). In Fig.~\ref{fig:dens} we show the radial density profile 
for both the clusters which is calculated using successive annuli of 
50 pixels ($\sim 35\arcsec$) around the cluster centre. For NGC 637, it 
flattens around $r=$ 250\arcsec and begin to merge with the background stellar 
density. Therefore, we consider 250\arcsec as the cluster radius. A smooth 
dashed line represents a fit by \citet{king62} profile:

   \begin{equation}
     f(r) = \frac {f_0}{1+(r/r_{c})^2} + f_b
     \label{eqn:king}
   \end{equation}

\noindent
where $f_0$ is the central density, $r_c$ is the core radius of the cluster 
and $f_b$ is the background density. For NGC 637, these parameters 
are $f_0 = 10.96 \times 10^{-3}$ star/arcsec$^2$, $r_c=39\farcs6$ and 
$f_b = 6.15 \times 10^{-3}$ star/arcsec$^2$. In case of NGC 957, 
\citet{king62} profile fitting could not converge because of large errors 
in the value of $f(r)$. Radius of this cluster is assumed as 260\arcsec 
considering the field star density on the basis of last two points 
in the radial density profile. For NGC 637 and NGC 957, the radius has 
been listed as 100\arcsec and 300\arcsec in \citet{dias02} catalog. Present 
estimate of the radius is larger for the cluster NGC 637 while it is smaller 
for NGC 957.

  \begin{figure*}
    \centering
    \includegraphics[width=18cm]{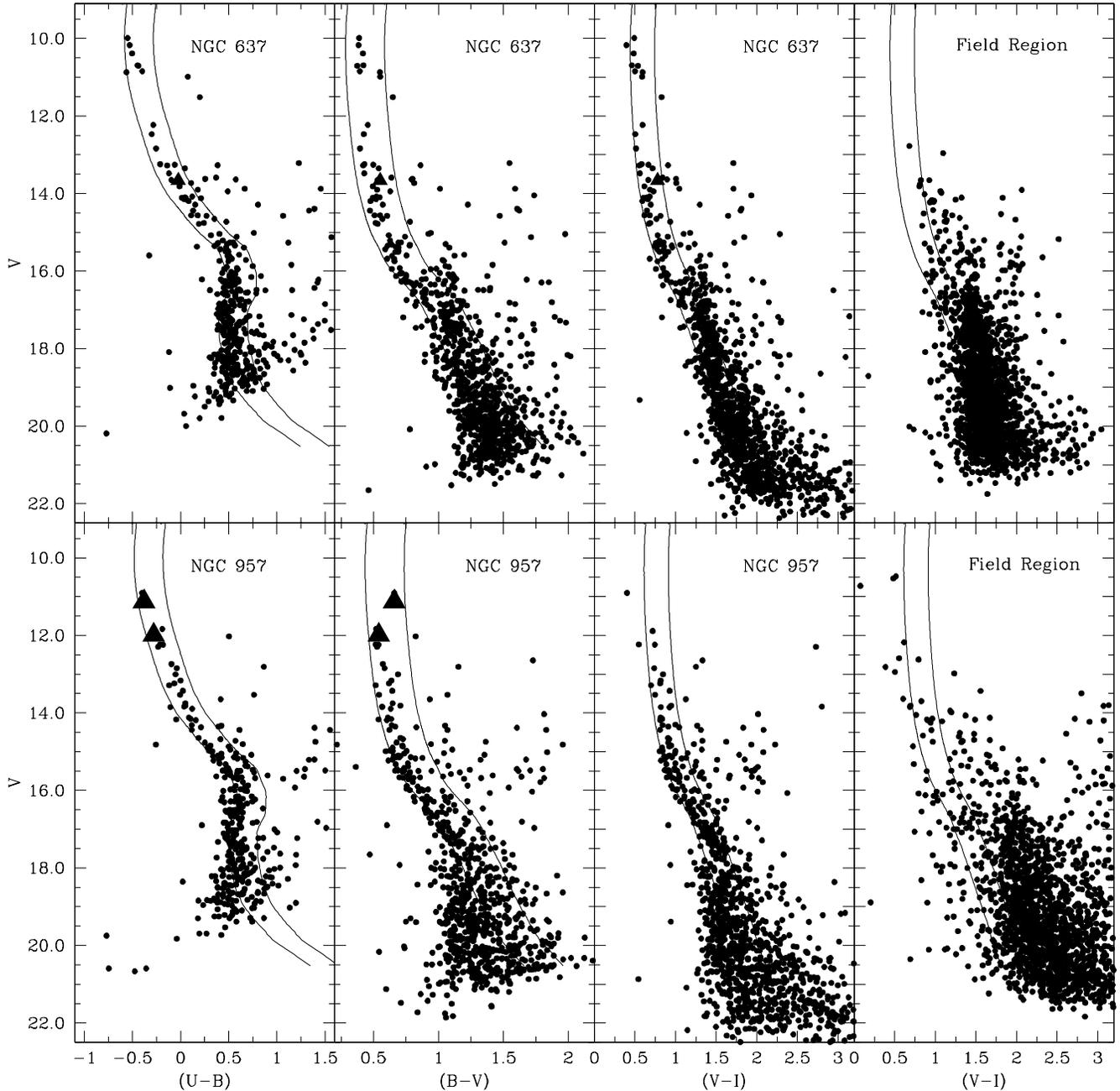}
    \caption{ The $V$, $(U-B)$; $V$, $(B-V)$ and $V$, $(V-I)$ CMDs for 
      the cluster NGC 637 and NGC 957 using stars within the cluster radius 
      and $V$, $(V-I)$ CMDs for the corresponding field regions. 
      Black triangles show the emission stars observed by us. Solid lines 
      represent the blue and red envelope of the cluster MS.} 
    \label{fig:cmd}
  \end{figure*}

\subsection{Colour-magnitude diagrams} \label{sec:cmd}

The $V, (U-B); V, (B-V)$ and $V, (V-I)$ CMDs of NGC 637 and NGC 957 alongwith 
the $V, (V-I)$ CMD of the corresponding field regions are shown in Fig.~\ref{fig:cmd}. 
To get the clear sequence of the cluster we plot only the stars within 
the cluster radius. It is evident that the morphology of CMDs for NGC 637 and 
NGC 957 is typical of a young age open star clusters. The CMDs extends 
down to $V \sim$ 21.0 mag except in $V$, $(U-B)$ CMDs where it is only up 
to $V \sim$ 19 mag. Main-sequence (MS) of the clusters is clearly visible upto 
$V\sim$ 17.0 mag. The MS fainter than 17.0 mag has more scatter and field 
star contamination is also more evident for fainter stars and because of this 
it is hard to delineate the field stars from the cluster members, only on the 
basis of their closeness to the main populated area of the CMD as field 
stars at cluster distance and reddening also occupy the same area. Proper motion 
or radial velocity data are very useful in separating cluster members from 
the field stars. But due to unavailability of this kind of data in the 
literature we use photometric criterion to separate probable cluster members 
from the field stars. We selected members by defining the blue and red 
envelope around the MS which is shown in the CMDs of both the clusters. A 
star is considered as a non-member if it lies outside the marked region in 
the CMDs. In Table~\ref{tab:freq}, we have listed the expected number of 
field stars using $V$, $(V-I)$ CMD of the field region. From this Table we 
can estimate the frequency distribution of stars in different parts of the 
CMD. It is also clear that all photometric probable members cannot be cluster 
members and non-members should be subtracted in the studies of cluster mass 
function etc. However, probable members located within a cluster radius 
can be used to determine the cluster parameters, as they have 
relatively less contamination due to field stars and the same has been assumed
in the subsequent analysis.

\setcounter{table}{3}
\begin{table*}
\centering
\caption{Frequency distribution of the stars in the $V$, $(V-I)$
diagram of the cluster and field regions. $N_{B}$, $N_{S}$ and $N_{R}$ denote
the number of stars in a magnitude bin blueward, along and redward of the
cluster sequence respectively. The number of stars in the field regions are 
corrected for area differences. $N_{C}$ (difference between the $N_{S}$ value
of cluster and field regions) denotes the statistically expected number of
cluster members in the corresponding magnitude bin.}

\begin{tabular}{c|ccc|ccc|c|c|ccc|ccc|c}
\hline
&&&&NGC 637&&&&&&&NGC 957&&\\
\cline{2-7} \cline{9-15}
$V$ range &\multicolumn{3}{|c|}{Cluster region} & \multicolumn{3}{|c|}{Field region}& &\multicolumn{3}{|c|}{Cluster region} & \multicolumn{3}{|c|}{Field region}& \\
&$N_{B}$&$N_{S}$&$N_{R}$&$N_{B}$&$N_{S}$&$N_{R}$&$N_{C}$ &$N_{B}$&$N_{S}$&$N_{R}$&$N_{B}$&$N_{S}$&$N_{R}$&$N_{C}$\\
\hline
 12 - 13 & 0 & 3 & 0 & 1 & 0 &  0& 3   & 0 & 2 & 2 & 0 & 0 &  2& 2  \\
 13 - 14 &10 & 8 & 0 & 5 & 0 &  0& 8   & 0 & 7 & 4 & 0 & 0 &  5& 7  \\
 14 - 15 & 6 &15 & 0 &10 & 0 &  0& 15  & 0 &16 &11 & 0 & 0 & 11& 16 \\
 15 - 16 &17 &22 & 0 &17 & 3 &  0& 19  & 0 &35 &27 & 0 & 0 & 21& 35 \\
 16 - 17 &40 &42 &1  &28 & 13&  0& 29  & 1 &34 &20 & 0 & 7 & 38& 27 \\
 17 - 18 &48 &62 &2  &55 & 24&  2& 38  & 3 &45 &31 & 2 & 7 & 80& 38 \\
 18 - 19 &48 &95 &4  &52 & 70&  5& 25  &17 &66 &48 & 3 & 26&107& 40 \\
\hline
\end{tabular}
\label{tab:freq}
\end{table*}

\subsection{Colour-colour diagram}

To estimate the interstellar extinction towards the clusters we plot 
$(U-B)$ versus $(B-V)$ diagrams using probable cluster members. The intrinsic 
zero-age main-sequence (ZAMS) given by \citet{schmidtkaler82} is fitted
by the dotted curve assuming the slope of reddening $E(U-B)/E(B-V)$ as 0.72. 
The ZAMS fitted to the MS stars of spectral type earlier than A0 provides 
a mean value of $E(B-V)=0.64\pm0.05$ mag for NGC 637 and $E(B-V)=0.71\pm0.05$ 
for NGC 957. Our derived values of reddening agree fairly well with the 
values estimated by others (\S\ref{sec:intro}).

   \begin{figure}
     \centering
     \hbox{
     \includegraphics[width=9cm]{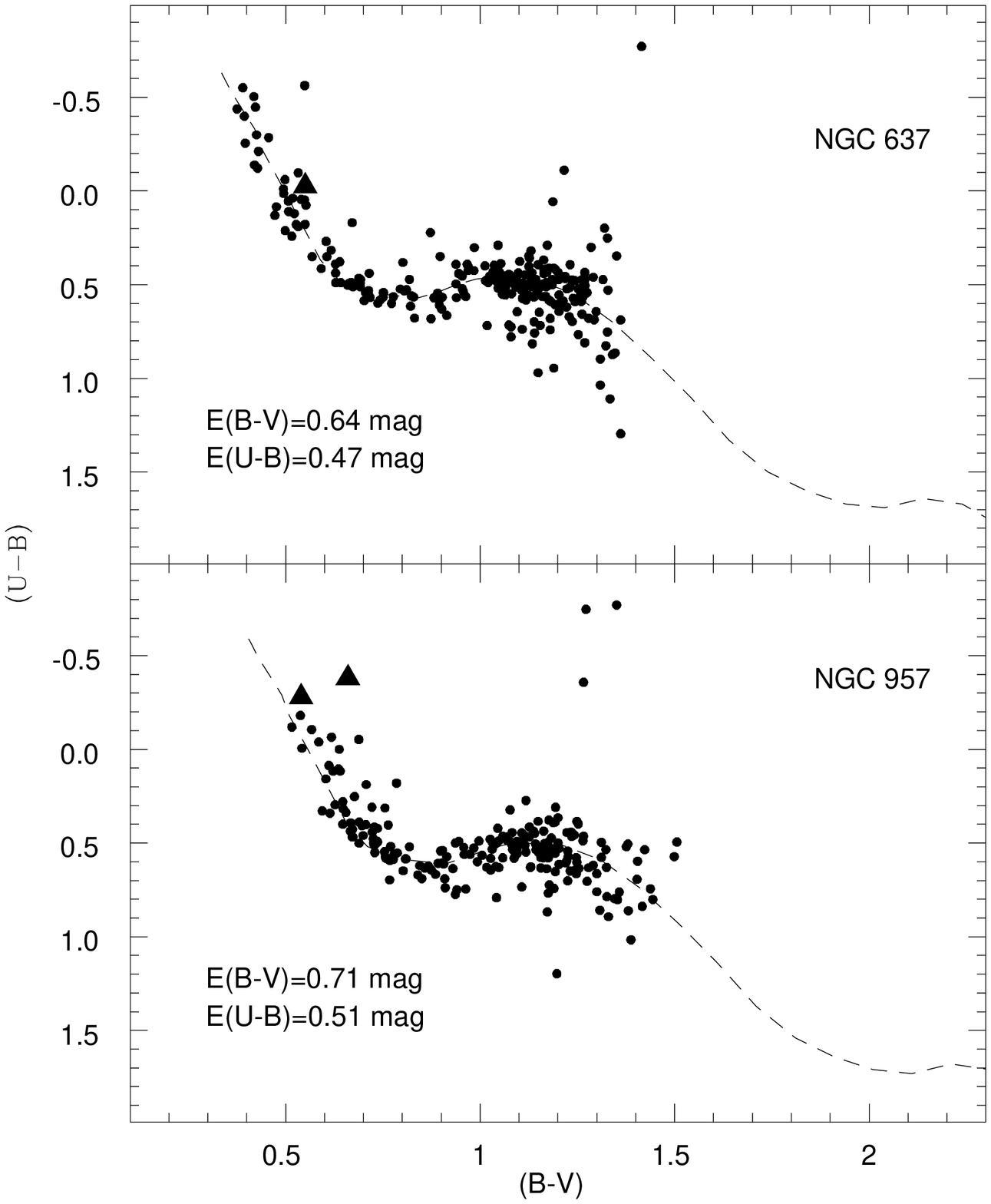}
     }
     \caption{ The $(U-B)$ versus $(B-V)$ colour-colour diagram of the 
           clusters. The continuous curve represents locus of 
           Schmidt-Kaler's (1982) ZAMS for solar metallicity. The filled 
           triangles are the emission type stars observed by us.} 
     \label{fig:cc}
     \end{figure}

  \begin{figure}
    \centering
    \includegraphics[width=9cm]{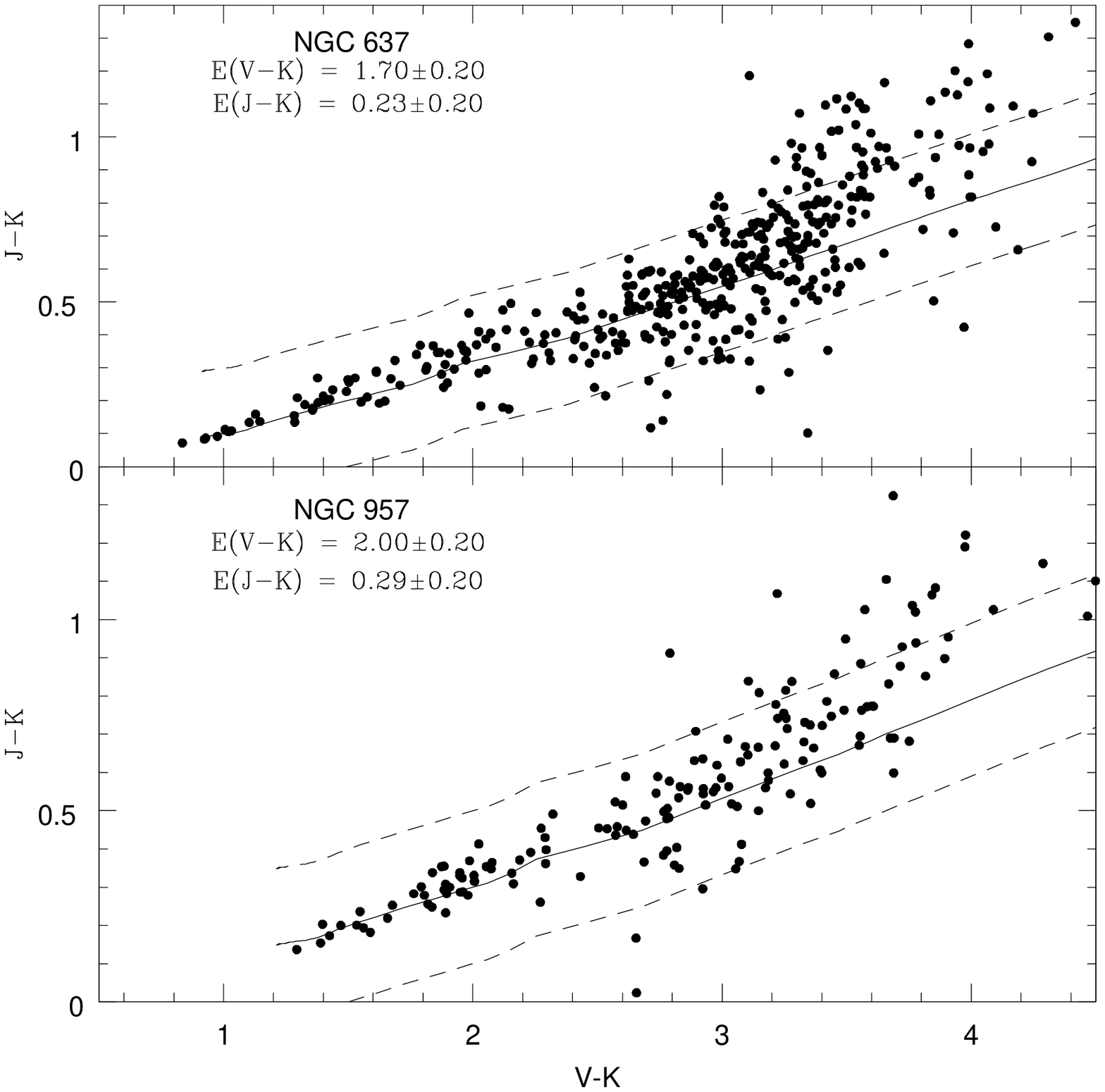}
    \caption{The plot of $(J-K)$ versus $(V-K)$ colour-colour diagram of the 
cluster for the stars within the cluster radius. The solid line is the ZAMS 
of $Z =$ 0.02 fitted for the marked values of the excesses while short dash 
lines show the errorbars.} 
  \label{fig:jhk_ir}
  \end{figure}

\subsection{Interstellar extinction in near-IR} \label{sec:extir}

2MASS $JHK$ data is used to study the interstellar extinction in 
combination with the optical data.  The $K_s$ magnitude are converted into $K$ 
magnitude following \citet{persson98}. The $(J-K)$ versus $(V-K)$ diagrams 
for both the clusters are shown in Fig. 7. The ZAMS taken 
from \citet{schaller92} for $Z=$ 0.02 is shown by solid line. The fit of 
ZAMS provides $E(J-K) = 0.23\pm0.20$ mag and $E(V-K) = 1.70\pm0.20$ mag 
for NGC 637 and $E(J-K) = 0.29\pm0.20$ mag and $E(V-K) = 2.00\pm0.20$ mag 
for NGC 957. The ratios $\frac{E(J-K)}{E(V-K)} \sim 0.14\pm0.30$ 
and $0.15\pm0.20$ for NGC 637 and NGC 957 respectively are in good 
agreement with the normal interstellar extinction value of 0.19 
\citep{cardelli89}. However, scattering is more due to large error in 
$JHK$ data.

\subsection{Extinction Law}

To investigate interstellar extinction law towards the clusters under 
study, we selected stars with spectral type earlier than A0. These 
have been identified from 
their location in apparent colour-colour ($U-B$ versus $B-V$) and CM 
diagrams, having $V<$ 15 mag, and $(B-V)<$ 0.70 mag, and for these stars, 
the MS spectral type have been obtained using $UBV$ photometric 
Q-method \citep{johnson53}. The colour excesses are determined by 
subtracting intrinsic colours from observed colours. The intrinsic 
colours are derived from the colour relation given by \citet{fitzgerald70} 
for $(U-B)$ and $(B-V)$; by \citet{johnson66} for $(V-R)$ and $(V-I)$ and
by \citet{koornneef83} for $(V-J)$, $(V-H)$ and $(V-K)$. To get the intrinsic 
colors for $V-R$ and $V-I$, the present Cousins $RI$ system are converted to 
the Johnson $RI$ using the relation given by \citet{bessell79}. 
Table~\ref{tab:ratio} lists the colour excess ratios estimated for both 
the clusters alongwith the ratios for normal interstellar matter 
(\citet{cardelli89}. The colour excess ratios generally agree 
within 3$\sigma$ with those given for the normal interstellar extinction law. 
This indicates that the interstellar extinction law is normal in the 
direction of both the clusters. 

Using the relation $R_V = 1.1 E(V-K)/E(B-V)$ \citep{whittet80}, we 
derived the values of $R_V$ as 3.1 for NGC 637 and 3.2 for NGC 957 and this 
indicates normal interstellar extinction law in the direction of both the 
clusters.

\begin{table*}
\caption{A comparison of extinction law in the direction of cluster with normal
extinction law given by \citet{cardelli89}.}
\begin{center}
\begin{tabular}{cccccccc}
\hline
&&&&&&&\\
Objects&$\frac{E(U-B)}{E(V-J)}$&$\frac{E(B-V)}{E(V-J)}$&$\frac{E(V-R)}{E(V-J)}$&$
\frac{E(V-I)}{E(V-J)}$&$\frac{E(V-H)}{E(V-J)}$&$\frac{E(V-K)}{E(V-J)}$&$\frac{E(
J-K)}{E(V-K)}$\\
&&&&&&&\\
\hline
Normal value&0.32&0.43&0.35&0.70&1.13&1.21&0.19\\
NGC 637&0.32$\pm$0.03&0.45$\pm$0.04&0.36$\pm$0.03&0.69$\pm$0.05&1.11
$\pm$0.08&1.19$\pm$0.07&0.15$\pm$0.30\\
NGC 957&0.34$\pm$0.02&0.47$\pm$0.01&0.39$\pm$0.03&0.75$\pm$0.05&1.12
$\pm$0.03&1.19$\pm$0.02&0.14$\pm$0.30\\
\hline
\end{tabular}
\end{center}
\label{tab:ratio}
\end{table*}

\subsection{Distance to the clusters}

  \begin{figure*}
    \centering
    \includegraphics[width=18cm]{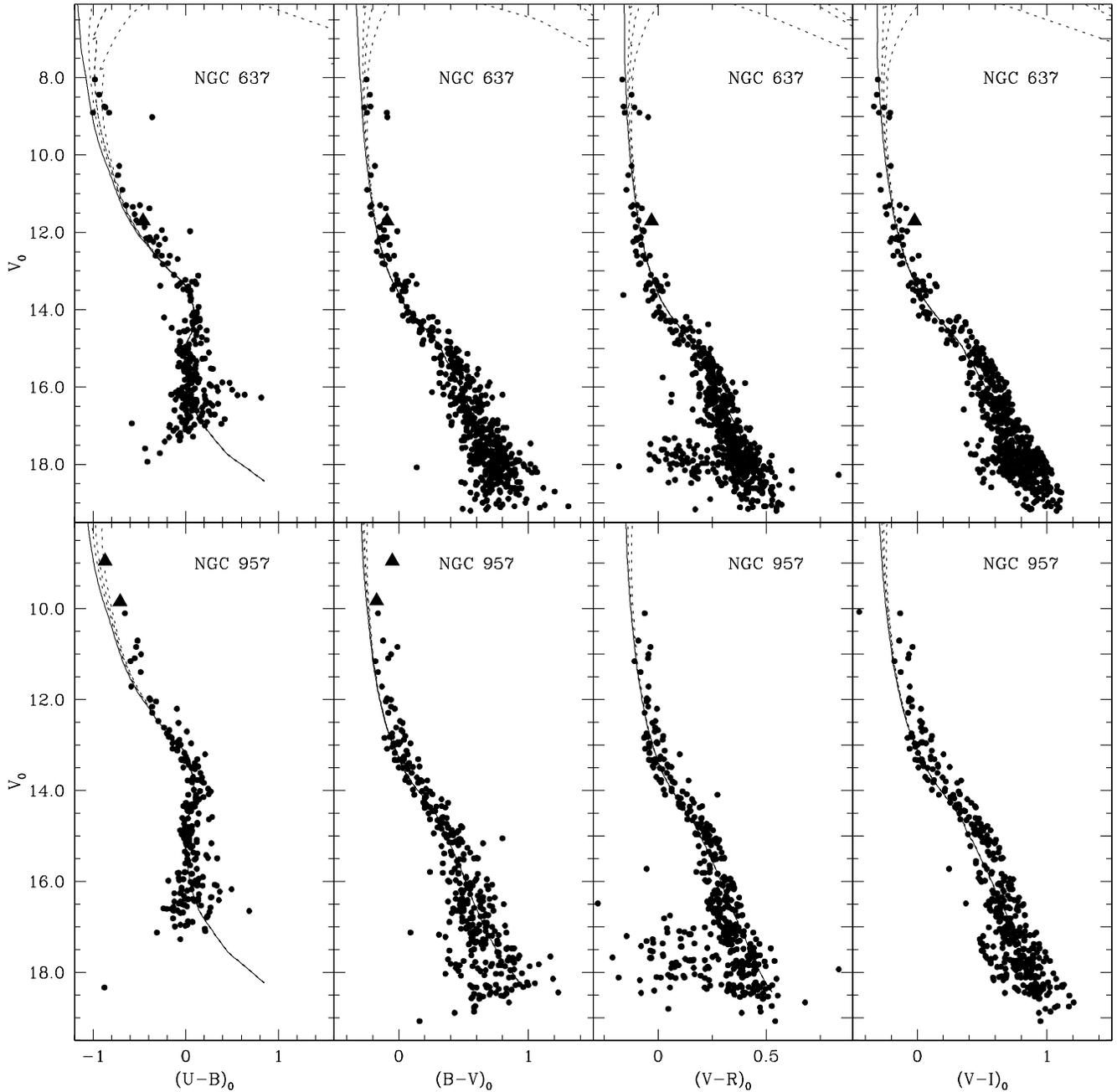}
    \caption{ The intrinsic colour-magnitude diagram of the clusters. The 
       continuous solid lines curve are the ZAMS given by \citet{schaller92} 
       for $Z =$ 0.02 and the dotted lines are isochrones of 
       log(age) $=$ 6.8,7.0 and 7.2 (shown left to right respectively) 
       for NGC 637 and NGC 957.}
    \label{fig:dist}
  \end{figure*}

   \begin{figure*}
     \centering
     \hbox{
     \includegraphics[width=9cm]{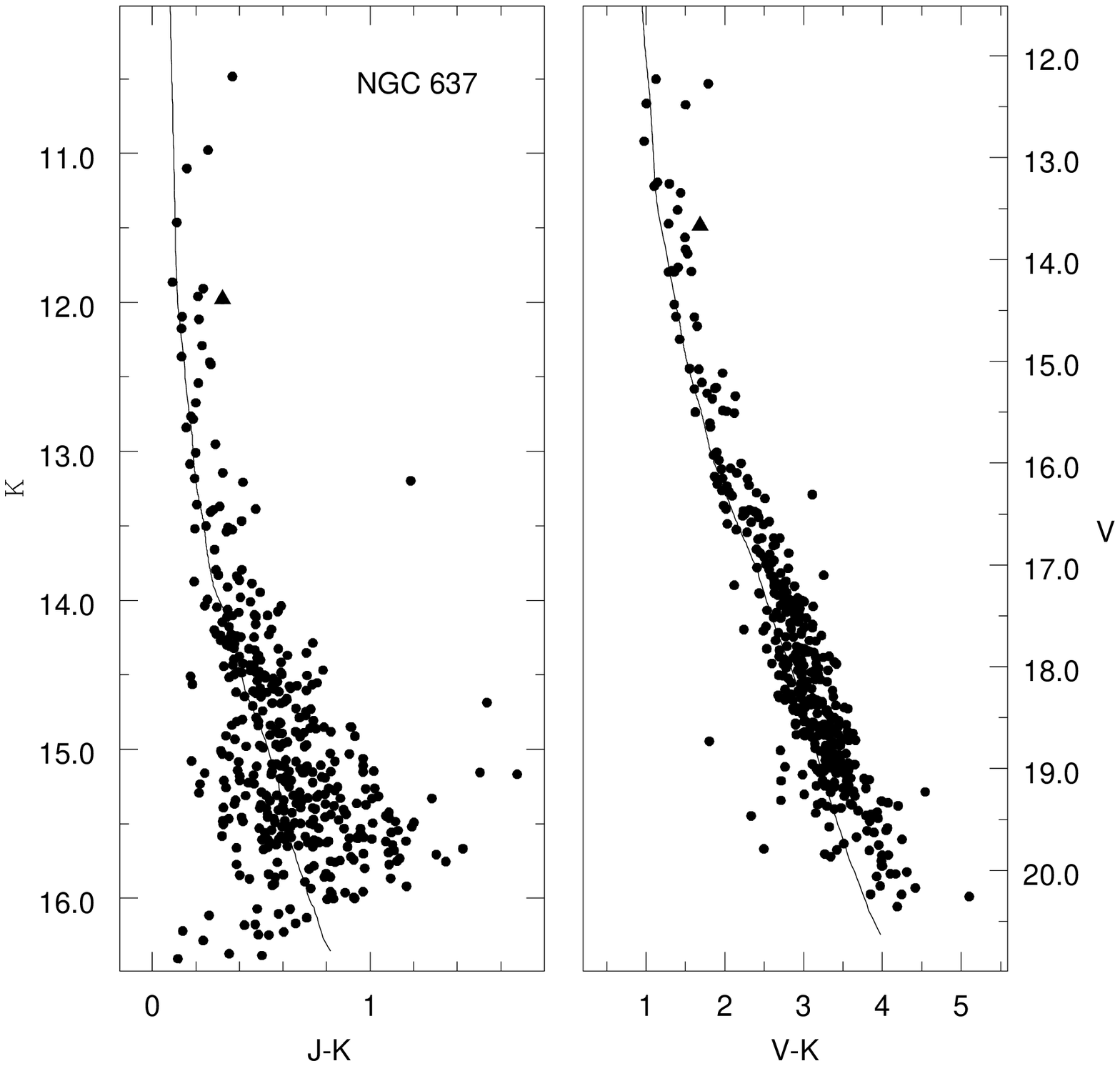}
     \includegraphics[width=9cm]{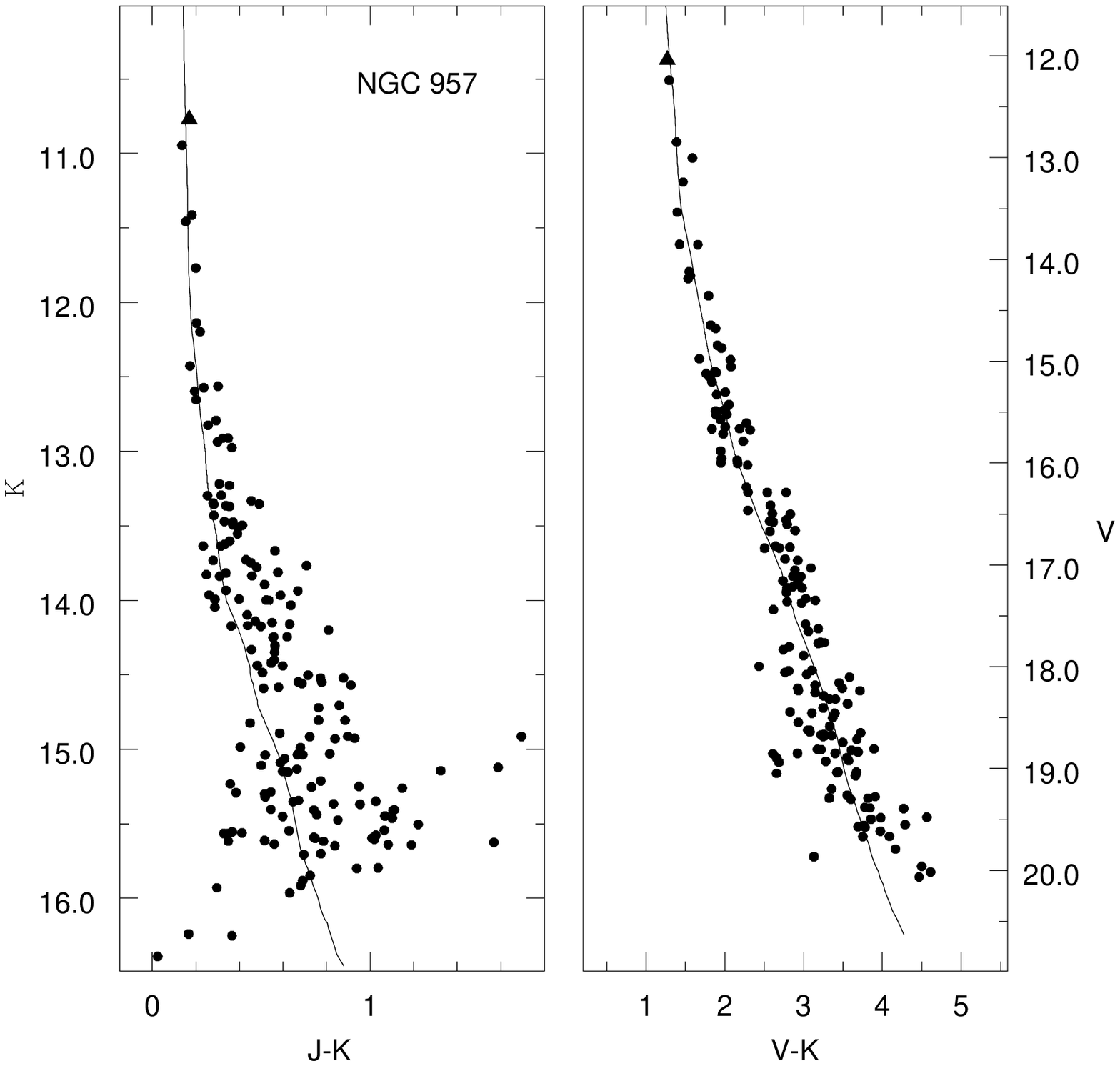}
     }
     \caption{The $K$ versus $(J-K)$ and $V$ versus $(V-K)$ CMDs of 
         the clusters using probable cluster members. The solid curve 
         represents the isochrones of log(age) $=$ 7.0 for NGC 637 and NGC 957 
         taken from \citet{schaller92} for $Z =$ 0.02.}
     \label{fig:cmdir}
   \end{figure*}

The ZAMS fitting procedure was used to derive distances to the clusters. 
Fig.~\ref{fig:dist} shows the intrinsic CMDs of probable cluster members 
as selected in \S\ref{sec:cmd} for the clusters NGC 637 and NGC 957. 
Employing mean $E(B-V)$ for the clusters, we converted 
apparent magnitude and colours into the intrinsic one using the method
as described in \citet{yadav02}. A theoretical ZAMS \citep{schaller92} 
for $Z =$ 0.02 is visually fitted to the blue part of intrinsic 
CMDs in $V_{0}$, $(U-B)_0$; $V_0$, $(B-V)_0$; $V_0$, $(V-R)_0$ and $V_0$, 
$(V-I)_0$. This gives an intrinsic distance modulus, $(m-M)_0$, of 
$12.0\pm0.2$ mag for NGC 637 and $11.7\pm0.2$ mag for NGC 957. The fact 
that we were able to find faint probable cluster members allow us to get 
a better definition of the cluster lower main sequence which in turn improves 
the estimation of the distances. The above distance moduli corresponds to a 
distance of $2.5\pm0.2$ kpc for NGC 637 and 
$2.2\pm0.2$ kpc for NGC 957. For NGC 637, it is similar to the value 2.5 kpc 
derived by \citet{huestamendia91} while it is less than the value 2.75 kpc 
derived by \citet{phelps94}. In case of NGC 957, our derived values are 
similar to the values given in literature (\S\ref{sec:intro}).

\subsection{Age of the clusters}

Age of the cluster is determined by visually fitting the theoretical stellar 
evolutionary isochrones \citep{schaller92} for $Z=0.02$ with the 
observed intrinsic CMDs of the clusters as shown in 
Fig.~\ref{fig:dist}. We chose to use the isochrones of $Z=0.02$, as the 
present target clusters are young and also many recent studies, for example, 
by \citet{twarog97}, does indicate that between 6.5 to 10 kpc galactocentric 
zone, the metallicity shows no gradient and its value is consistent with 
the solar value within the uncertainties. Also for cluster 
NGC 637, \citet{phelps94} considered solar metallicity.

Isochrones of log(age) $=$ 6.8, 7.0 and 7.2 are shown with dotted lines from 
left to right in Fig.~\ref{fig:dist} and these visually fitted isochrones 
to the brighter stars indicate that CMDs are represented well with an isochrone
of about 10 Myr for both the clusters. Furthermore, the isochrones of 
log(age) $=$ 6.8 and 7.2 Myr seem to confine blue and red end of the upper 
main sequence stars. The most massive stars in NGC 637 and NGC 957 is about 
13 M$_{\odot}$ and it sets an upper limit of about 15 Myr for the MS age of 
the clusters. We have therefore adopted a mean age of 10 Myr with an 
uncertainty of about 5 Myr for both the clusters. 

The present age estimate for NGC 637 is not too different from the values of 
15 Myr as derived by \citet{huestamendia91} and 4 Myr by \citet{phelps94}. 
For NGC 957, \citet{gerasimenko91} has given an age of 5 Myr, 
which agrees within error with the present estimate.
 
Using optical and near-IR data we re-determined distance and age of both the 
clusters. We plot $V$ versus $(V-K)$ and $K$ versus $(J-K)$ CMDs in 
Fig~\ref{fig:cmdir}. The theoretical isochrones given by \citet{schaller92}
for $Z =$ 0.02 and log(age) $=$ 7.0 have been overplotted in the CMDs of 
NGC 637 and NGC 957. The apparent distance moduli $(m - M)_{V, (V-K)}$ 
and $(m-M)_{K, (J-K)}$ turn out to be 14.0$\pm$0.3 and 12.0$\pm$0.3 mag 
for the cluster NGC 637 and 14.0$\pm$0.3  and 12.1$\pm$0.3 mag for the 
cluster NGC 957. Using the reddening values estimated in \S\ref{sec:extir}, 
we derived a distance of 2.5$\pm$0.3 for NGC 637 and 2.2$\pm$0.3 kpc for 
NGC 957. Both age and distance determination for the clusters are thus 
in agreement with the estimates using optical data. However, 
scattering is larger due to the large errors in $JHK$ mags.

\subsection {Gaps in the main-sequences}

A peculiarity that appears in the intrinsic CMDs is the presence of gaps in 
the MS (see Fig.~\ref{fig:dist}). A MS gap is present between 
$9<V_0<10$ ($11<M/M_\odot<8$) in NGC 637. This gap has already been noticed 
by \citet{huestamendia91} and \citet{phelps93}. The deficiency of 
stars between $11<V_0<12$ ($4<M/M_\odot<3$) is also seen in NGC 957. To check 
the significance of the gap, we computed the probability that a lack of 
stars in the mass interval is a result of random processes 
\citep[see][]{scalo86,giorgi02} by using the relation 
$P_{gap} = (M_{sup}/M_{inf})^{(-N\times x)}$, where $M_{sup}$ and $M_{inf}$ 
are the upper and lower masses of the gap and $x$ is the exponent of mass 
function while $N$ is the number of cluster members located above the gap. 
Adopting the MF slope in advance (\S\ref{sec:mfu}), we find $P_{gap} = 4\%$ for 
NGC 637 and 16\% for NGC 957. The low resulting probabilities for both the 
clusters suggest that gaps in the MS are not due to the random processes. 
Hence, observed gaps in the MS of the clusters are real features, though we 
note that the significance test of gaps in MS are only approximate because 
the terms of the test were set after the data were observed. Specifically, 
upper and lower masses of the gap were derived after seeing the data. Thus, 
the quantitative statistical significance of the gaps has been overestimated.  

There are many reports about the MS gaps of other star clusters in the 
literature \citep{mermilliod76,forbes96,herbst82,baume04,kumar08}. 
A detailed description of gaps 
in several open clusters has recently been given by \citet{rachford00}. 
The explanation of the gaps found in open clusters range from the 
onset of convection in the stellar envelopes \citep{bohmvitense74} 
to peculiarities in the Balmer jump and Balmer lines as suggested 
by \citet{mermilliod76}. \citep{ulrich71a,ulrich71b} speculated that gaps in 
young clusters might be produced by the presence of ${}^{3}$He isotopes that 
halt gravitational contraction 1-2 mags above the main-sequence. 
Mermilliod's gaps occur in the range of B-type stars while B\"{o}hm-Vitense \& 
Canterna's gaps are found at less massive stars. Observed gap in NGC 637 
may be related to that found by \citet{mermilliod76} at the types B1-B2.

   \begin{figure}
     \centering
     \hbox{
     \includegraphics[width=9cm]{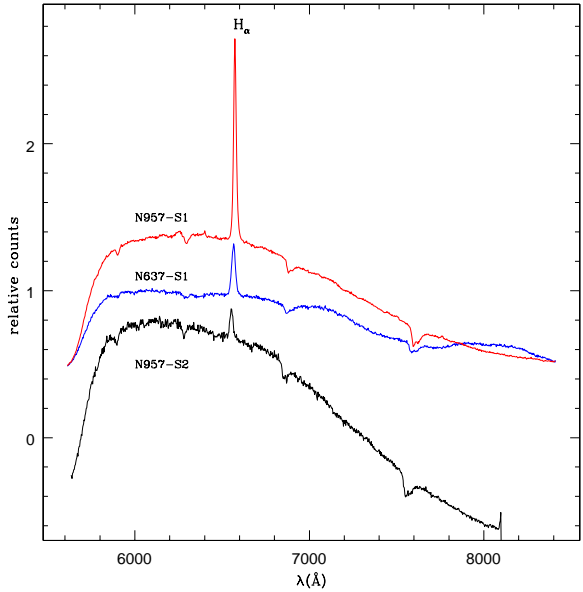}
     }
     \caption{Slitless spectra of three emission line stars in NGC 637 and NGC 957} 
     \label{fig:sp}
   \end{figure}

   \begin{figure}
     \centering
     \includegraphics[width=9cm]{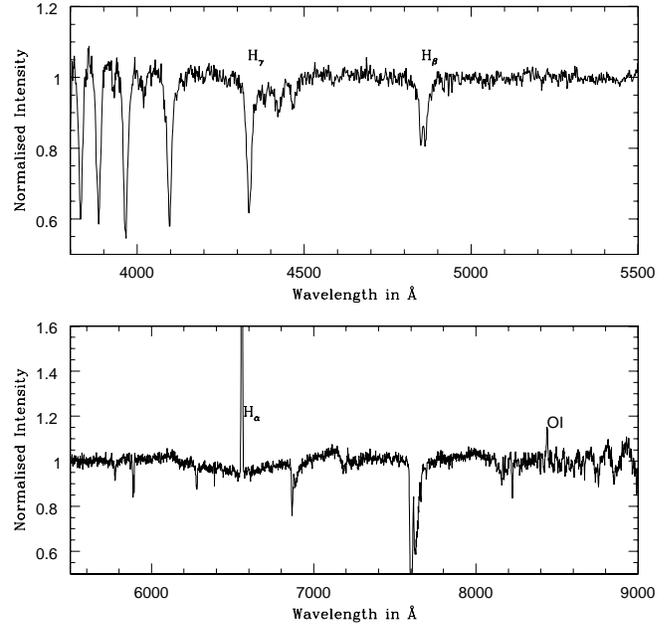}
     \caption {The normalised, continuum fitted spectra of the emission star 
          NGC 637-S1 in the wavelength range 3700-9000 \AA{} in the cluster 
          NGC 637.}
     \label{fig:spec637}
   \end{figure}

   \begin{figure}
     \centering
     \includegraphics[width=9cm,height=9cm]{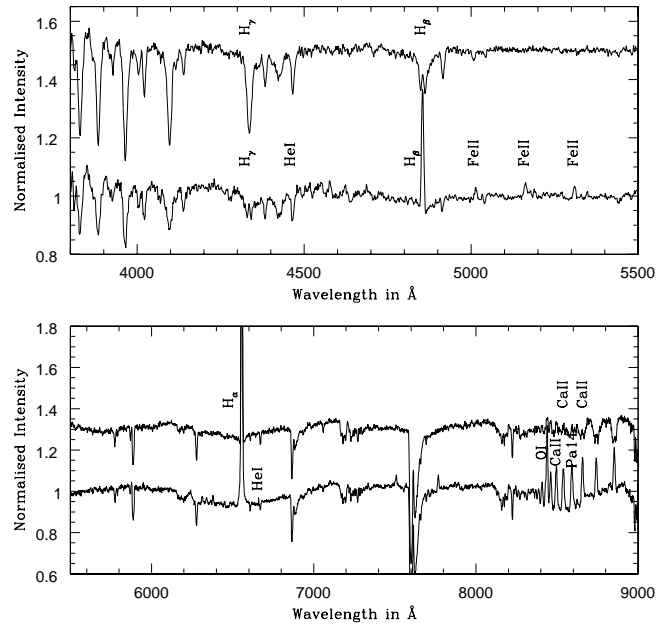}
     \caption {The normalised, continuum fitted spectra of the emission stars 
         in NGC 957 in the wavelength range 3700-9000 \AA{}. The top spectra 
         corresponds to star NGC 957-S2 and the bottom one corresponds to star 
         NGC 957-S1.} 
     \label{fig:spec957}
   \end{figure}

\subsection{Emission type stars}

The Be star phenomenon in young open clusters is  known to arise due to
mass loss from early type stars during its evolution on the MS and
are found to be maximum for clusters with 10-25 Myr age \citep{fabregat00}. 
About 20\% of the early-type ($<$ B5) stars show Be 
phenomenon in young clusters of our Galaxy \citep{maeder99}. However, 
in some young clusters, resulting from on-going/episodic star formation, an 
over-abundance of Be stars ($> 30\%$) are caused due to presence of Herbig Be 
stars, which are usually associated with nebulosity and show large 
near-infrared excess \citep{subramaniam05,subramaniam06}.

In the slit-less spectra, the emission-line stars are usually seen as 
continuum enhancement and they were identified by visual inspection of each 
spectra. We could identify emission-line stars with a typical limit of about 
$V \sim$ 16 mag in 10-min exposure. This roughly corresponds to a late
F-type star for present clusters. The positions of the emission stars are 
identified from the  $R$-band images and the spectra are extracted from the 
slit-less spectral images. The spectra of individual stars are extracted with 
an aperture width of about 5 pixels 
\citep[see][for detailed description]{subramaniam05,subramaniam06}. 
We identified an emission line star in NGC 637 for the 
first time. We also identified two emission stars in NGC 957, which are 
previously known. The slitless spectra of one star in NGC 637 and two stars 
in NGC 957 are shown in Fig. 10. These spectra 
cannot be used for quantitative analysis, therefore, we obtained the slit 
spectra of the identified emission stars. The slit spectra of NGC 637-S1 
is shown in Fig.~\ref{fig:spec637}. The spectra show H$_\alpha$ in emission, 
H$_\beta$ in emission within absorption. Two emission stars in NGC 957 were 
also observed in the slit mode, the spectra are shown in Fig.~\ref{fig:spec957}.
NGC 957-S1 is found to show H$_\alpha$, H$_\beta$ and H$_\gamma$,
along with a large number of FeII, OI and Paschen lines in emission. 
NGC 957-S2 shows partially filled up H$_\beta$, faint emissions in Paschen 
lines along with H$_\alpha$ emission. Photometric data with position of all 
three emission stars identified in the present study are listed in
Table 6. Because of the CCD saturation in the 
present observations, photometric data of two emission stars in NGC 957 are 
taken from \citet{hoag61} and tabulated in Table 6. Thus, in the
CMDs, the brighter and redder star is S1 and the fainter and bluer star is S2.
The values of H$_\alpha$ equivalent width, stellar rotation, vsini, as 
estimated from the width of the HeI (4471\AA) line and the disk rotation, as 
estimated from the width of the H$_\alpha$ profile are also tabulated.

The spectral type of the star is estimated from its $V$ and $(B-V)$ magnitudes 
and using the estimated cluster reddening and distance. In NGC 637, the 
emission star is found to be a late B-type (B6.5V; NGC 637-S1) star with low 
infrared excess. This star is located well below the turn-off. The stellar 
vsini is found to be very large, 450\kms{} and disk vsini is found to be 
much smaller, 180\kms. Thus, this is a Classical Be star in the
MS evolutionary phase and with large stellar rotation. Thus, this young 
clusters houses a fast rotating Classical Be star.

The stars NGC 957-S1 and NGC 957-S2 in the cluster NGC 957 were earlier 
identified by \citet{kohoutek99} as emission type stars. The emission 
stars are found to belong to early B-type (S1 - B2V and S2 - B2.5V) 
from photometry. These stars are located near the tip of the cluster MS, but 
below the turn-off. These stars have stellar vsini more than 300\kms{}  
indicating these are Classical Be candidates. These 
stars also have low near-infrared excess which supports the above.
The star with relatively large H$_\alpha$ emission, S1, is found to have a 
relatively slow rotating disk. Thus, this is another young cluster which 
houses Classical Be stars.

\begin{table*}
\centering
\caption {Photometric information about emission stars. Photometric data 
of NGC 957-S1 and NGC 957-S2 star are taken from Hoag et al. (1961).}

\hspace{2cm}

\scriptsize
\begin{tabular}{|c|l|c|c|c|c|c|c|c|c|}
\hline
Objects&Coordinate (J2000)&$V$&$(U-B)$&$(B-V)$&$(V-R)$&$(V-I)$ &H$_\alpha$ EW & vsini(disk)& vsini(star)\\ \hline
NGC 637-S1&$\alpha=01^{h}43^{m}23^{s}$.1&13.67&$-$0.03&0.55&0.38&0.78 & -21.97 & 179.65 & 450.37 \\
          &$\delta=64^{0}01^{\prime}19^{\prime\prime}$.2&&&&&&&&\\
NGC 957-S1&$\alpha=02^{h}33^{m}08^{s}$.2&11.13&$-$0.38&0.66&& & -37.99& 184.32& 342.72\\
          &$\delta=57^{0}28^{\prime}10^{\prime\prime}$.8&&&&&&&&\\
NGC 957-S2&$\alpha=02^{h}33^{m}10^{s}$.5&11.99&$-$0.28&0.54&& &-16.96& 251.62 & 308.85\\
          &$\delta=57^{0}32^{\prime}53^{\prime\prime}$.9&&&&&&&&\\
\hline
\end{tabular}
\label{emis_star}
\end{table*}


\section{Luminosity and Mass function} \label{sec:mfu}

To construct the luminosity function for the cluster, we used the deepest $V$ 
versus $(V-I)$ CMD. Photometric criteria (see \S\ref{sec:cmd}) have been used 
to select the cluster members. The same envelope as made for selection of 
cluster members is also drawn in the $V$ versus $(V-I)$ CMD of the 
field region (see Fig~\ref{fig:cmd}). Stars are counted within this 
envelope, for both cluster and field region. Difference between the 
counts in two fields after accounting for the difference in area between 
the cluster and field regions will be the observed cluster luminosity 
function. To get the correct luminosity function it is very essential to 
derive completeness factor of CCD data. To know about the completeness of our 
photometric data, we performed experiments with artificial stars using 
ADDSTAR routine in DAOPHOT II. Detailed description about the experiment 
is given in \citet{yadav02} and \citet{sagar98}. The completeness factors  
derived in this way are tabulated in Table~\ref{tab:cf} for NGC 637 and 
NGC 957 and it shows that the completeness factor is 
$\sim$ 93\% at 20 mag for both clusters. The completeness factor of 
field region is assumed to be 100\%.

   \begin{figure}
     \centering
     \includegraphics[width=9cm]{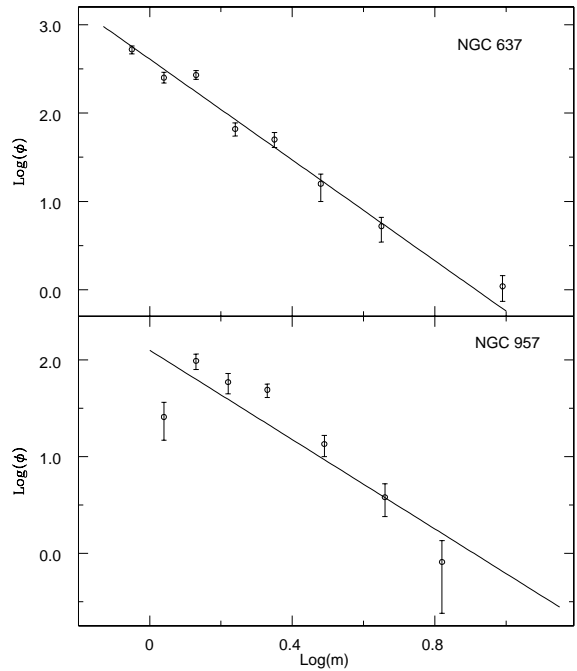}
     \caption{Mass function for NGC 637 and NGC 957 derived using 
          \citet{schaller92} isochrones.}
     \label{fig:mass}
   \end{figure}

Using the cluster's parameters derived in this analysis and theoretical models 
given by \citet{schaller92} we have converted LF to MF and the resulting 
MF is shown in Fig.~\ref{fig:mass}. The mass function slope can be derived by 
using the relation 
${\rm log}\frac{dN}{dM} = -(1+x) \times {\rm log}(M)+${}constant, 
where $dN$ represents the number of stars in a mass bin $dM$ with central 
mass $M$ and $x$ is the slope of MF. The slope of the MF is 
$x = 1.65\pm0.20$ and $1.31\pm0.50$ for the cluster NGC 637 and NGC 957 
respectively. Our derived values of mass function slope is in 
agreement within the error with the value 1.35 given by 
\citet{salpeter55} for field stars in solar neighbourhood. 

\setcounter{table}{6}
\begin{table}
\centering
\caption{Variation of completeness factor (CF) in the $V$, $(V-I)$ 
      diagram with the MS brightness.}
\begin{tabular}{|c|c|c}
\hline
$V$ mag range&NGC 637&NGC 957\\
\hline
12 - 13&0.99&0.99\\
13 - 14&0.99&0.99\\
14 - 15&0.99&0.99\\
15 - 16&0.98&0.98\\
16 - 17&0.97&0.98\\
17 - 18&0.95&0.94\\
18 - 19&0.94&0.94\\
19 - 20&0.93&0.93\\
\hline
\end{tabular}
\label{tab:cf}
\end{table}

There are many MF studies available in the literature using open star clusters 
in the Milky Way. Recently, \citet{piskunov04} studied 5 young open star 
clusters and found that stellar mass spectra of these clusters are well 
represented with a power law very similar to Salpeter value within the 
uncertainties. MF study of \citet{phelps93} using a sample of seven star 
clusters also conclude the Salpeter type MF slope in these clusters. 
These studies alongwith the others \citep{sanner01, sagar01, yadav02, yadav04} 
found their results consistent with the Salpeter value.


\section{Mass segregation} \label{sec:mse}

\begin{figure}
       \centering
       \includegraphics[width=9cm]{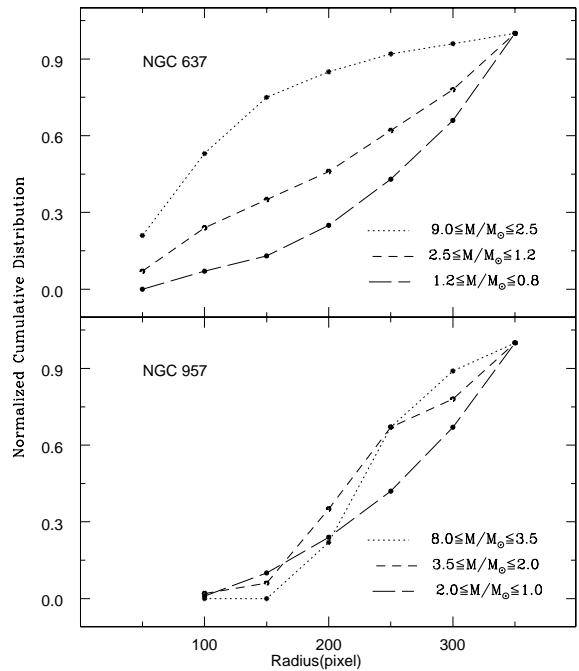}
       \caption{The cumulative radial distribution of stars in various mass 
                range.}
       \label{fig:mseg}
    \end{figure}

In order to study the mass segregation effect in the clusters, we 
plot the cumulative radial stellar distribution of stars for different masses 
in Fig~\ref{fig:mseg} and mass segregation effect is seen for both the 
clusters, meaning, higher-mass stars gradually sink towards the 
cluster center than the lower mass stars. Further, We performed K-S test to 
see the statistical significance of mass segregation. For NGC 637, we divided 
the MS stars in three magnitude range i.e. $12.0 \le V<15.0$, $15.0\le V<18.0$ 
and $18.0\le V<20.0$ corresponding to the mass range of 
$9.0\le M/M_{\odot}<2.5, ~2.5\le M/M_{\odot}<1.2$ and 
1.2$\le$ M/M$_{\odot}$$<$0.8. For NGC 957,
the magnitude range are $12.0\le V<14.0, 14.0\le V<16.0$ and 
$16.0\le V<18.0$ corresponding to the mass range of 
$8.0\le M/M_{\odot}<3.5$, ~$3.5\le M/M_{\odot}<2.0$ and 
$2.0\le M/M_{\odot}<1.0$.
The K-S test shows the mass segregation effect at confidence 
level of 99\% for NGC 637 and 70\% for NGC 957.

Mass segregation effect can be due to dynamical evolution or imprint of star 
formation or both. In the lifetime of star clusters, encounters between its 
member stars gradually lead to an increased degree of energy equipartition 
throughout the clusters. In this process the higher mass cluster 
members gradually sink towards the cluster center and transfer their kinetic 
energy to the more numerous lower-mass stellar component, thus leading to mass 
segregation. The time scale on which a cluster will have lost all traces of 
its initial conditions is well represented by its relaxation 
time $T_E$, which is given by 

\begin{equation}
T_{E} = \frac {8.9 \times 10^{5} N^{1/2} R_{h}^{3/2}}{ <m>^{1/2}log(0.4N)}
\end{equation}

\noindent
where $N$ is the number of cluster members, $R_{h}$ is the half-mass radius
of the cluster and $<m>$ is the mean mass of the cluster stars
\citep{spitzer71}. The value of $R_{h}$ has been assumed as half of 
the cluster radius derived by us. Using the above relation we estimated the 
dynamical relaxation time $T_E =$ 10 Myr for NGC 637 and NGC 957. This 
indicates that the age of both clusters is same as its relaxation age. 
Therefore, we conclude that both the clusters are dynamically relaxed.

\section{Conclusions} \label{sec:con}

We studied two open star clusters NGC 637 and NGC 957 using $UBVRI$ CCD 
and 2MASS $JHK$ data. Results obtained in the analysis are the following.

\begin{enumerate}

\item The radii of the clusters are obtained as 4\farcm2 and 4\farcm3 
which corresponds to 3.0 and 2.8 pc respectively at the distance of the 
clusters NGC 637 and NGC 957.

\item From the two colour diagram, we estimated 
$E(B-V) = 0.64\pm0.05$ mag for NGC 637 and $0.71\pm$0.05 mag for NGC 957. 
The $JHK$ data in combination with the optical data provide $E(J-K) = 
0.23\pm$0.20 mag and 0.23$\pm$0.20 mag while $E(V-K) = 1.70\pm$0.20 mag 
and $E(V-K) = 2.00\pm$0.20 mag for NGC 637 and NGC 957 respectively. Analysis 
indicates that interstellar extinction law is normal towards the direction of 
both the clusters.  

\item Distances to the cluster NGC 637 and NGC 957 are determined as 
2.5$\pm$0.2 and 2.2$\pm$0.2 kpc respectively. These distances are supported 
by the distance values derived using optical and near-IR data. 
An age of 10$\pm$5 Myr is determined for both NGC 637 and NGC 957 by 
comparing the isochrones of $Z =$ 0.02 given by \citet{schaller92}.

\item We discovered one emission type star in NGC 637. Two emission type stars 
in NGC 957 are also confirmed by the present study. These are classical Be 
star candidates. The two emission stars in NGC 957 are about to leave the MS while the emission star in NGC 637 is in the MS evolutionary phase. 
  
\item The mass function slopes $x = 1.65\pm0.20$ and $1.31\pm0.50$ are derived 
for NGC 637 and NGC 957 by considering the corrections of field star 
contamination and data incompleteness. Our analysis indicate that both the 
cluster are dynamically relaxed and one plausible reason 
of this relaxation may be the dynamical evolution of the clusters.

\end{enumerate}

{\bf ACKNOWLEDGMENTS}

We thank the anonymous referee for the useful comments and suggestions which 
surely helped to improve the paper. One of us (BK) acknowledge support from 
the Chilean center for Astrophysics FONDAP No. 15010003. This publication 
made use of data from the 
Two Micron All Sky Survey, which is a joint project of the University of 
Massachusetts and the Infrared Processing and Analysis Center/California 
Institute of Technology, funded by the National Aeronautics and Space 
Administration and the National Science Foundation. We are also much obliged 
for the use of the NASA Astrophysics Data System, of the Simbad database 
(Centre de Donn$\acute{e}$s Stellaires-Strasbourg, France) and of the WEBDA 
open cluster database.

\end{document}